\documentclass[aps,prd,twocolumn,showpacs,amsmath,amssymb,nofootinbib]{revtex4-1}
\usepackage{blindtext}
\usepackage{graphicx}
\usepackage{color}
\usepackage{times}
\usepackage{bm}
\usepackage{multirow}
\usepackage{float}
\usepackage{url}
\usepackage{natbib}
\usepackage{tensor}
\usepackage{bbm}
\usepackage[colorlinks=true,citecolor=blue,urlcolor=blue,linkcolor=red]{hyperref}
\usepackage[normalem]{ulem}
\usepackage{verbatim}
\usepackage{lipsum} 
\usepackage{tikz,xcolor,hyperref}
\usepackage{ifxetex,ifluatex}

\newcommand{\ep}{\varepsilon}

\def\Msun{\,M_{\odot}}
\def\MeV{\,\text{MeV}}
\def\fm3{\;\text{fm}^{-3}}
\def\MeVfm3{\,\text{MeV/fm}^{3}}
\def\km{\;\text{km}}

\newcommand{\apjl}{Astrophys. J. Lett.}

\newcommand{\aap}{Astron. Astrophys.}
\newcommand{\araa}{Annual Review of Astronomy \& Astrophysics}
\newcommand{\mnras}{Mon. Not. R. Astron. Soc.}

\newcommand{\nphysa}{Nuclear Physics A}
\newcommand{\physrep}{Physics Reports}

\DeclareUnicodeCharacter{2212}{-}

\definecolor{lime}{HTML}{A6CE39}
\DeclareRobustCommand{\orcidicon}{%
	\begin{tikzpicture}
	\draw[lime, fill=lime] (0,0) 
	circle [radius=0.17] 
	node[white] {{\fontfamily{qag}\selectfont \tiny ID}};
	\draw[white, fill=white] (-0.0625,0.095) 
	circle [radius=0.008];
	\end{tikzpicture}
	\hspace{-2mm}
}
\foreach \x in {A, ..., Z}{
	\expandafter\xdef\csname orcid\x\endcsname{\noexpand\href{https://orcid.org/\csname orcidauthor\x\endcsname}{\noexpand\orcidicon}}
}

\begin{document}
\title{Thermal x-ray studies of neutron stars and the equation of state}

\author{Zhiqiang Miao{\orcidA{}}$^{1,2}$}
\author{Liqiang Qi
$^{3}$}
\author{Juan Zhang
$^{3}$}
\author{Ang Li{\orcidB{}}$^{2}$}
\email{liang@xmu.edu.cn}
\author{Mingyu Ge
$^{3}$}
\email{gemy@ihep.ac.cn}

\affiliation{\it $^{1}$ Tsung-Dao Lee Institute, Shanghai Jiao Tong University, Shanghai 201210, China;\\
$^{2}$ Department of Astronomy, Xiamen University, Xiamen 361005, China;\\
$^{3}$ Key Laboratory of Particle Astrophysics, Institute of High Energy Physics, Chinese Academy of Sciences, Beijing 100049, China
}

\date{\today}

\begin{abstract} 
The understanding of neutron star equation of state hinges on a comprehensive analysis of multi-messenger, multi-wavelength data.
The recent scrutiny of PSR J0030+0451 data by NICER introduces complexities, unveiling a tension with another x-ray observation of the central compact object in HESS J1731-347, specifically concerning the mass-radius constraint of low-mass neutron stars. This tension persists when integrating NICER's updated data with LIGO/Virgo's gravitational-wave data from the GW170817 binary neutron star merger. 
Despite attempts to reconcile these disparate observations, the current combined data still can not distinguish different types of neutron stars -- whether they are pure neutron stars or hybrid stars. 
Bayesian inference indicates only modest changes in the posterior ranges of parameters related to the nuclear matter and deconfinement phase transition.
This ongoing exploration underscores the intricate challenges in precisely characterizing neutron stars. It 
also points out that it is possible to probe the equation of state at different density regimes from future more accurate radii of neutron stars with various masses.
\end{abstract}

\maketitle 

\section{Introduction}

The determination of the equation of state (EoS) for neutron stars (NSs) through quantum many-body computations proves to be a formidable task due to the complicated nonperturbative problems in physics~\citep{2020PhR...879....1H}.
Effective discrimination between different EoSs based on nuclear experiments and astrophysical observations is further hindered by a deficit and limited precision of measurements~\citep{2022EPJWC.26004001L}. 
Consequently, the study of EoS, namely how the pressure $P$ of NS matter system changes with the stellar (energy) density $\varepsilon$, relies on an iterative interplay between experiment and theory, with their reciprocal feedback proving crucial for progress~\citep{2018RPPh...81e6902B,2020JHEAp..28...19L,2021PrPNP.12003879B,2021ARNPS..71..433L,2021MNRAS.506.5916L}. 
Assuming that NSs obey general relativity, the EoS establishes a unique sequence of stars in hydrostatic equilibrium using the Tolman-Oppenheimer-Volkoff (TOV) equation. 
Accurate measurements of mass and radius of NSs can then be employed to constrain physical quantities in the EoS, describing the properties of dense nuclear matter (i.e.\ the incompressibility, the symmetry energy) and strangeness phase transitions (i.e.\ the critical chemical potential of the phase transitions, the sound velocity).

The direct measurement of the mass and radius of NSs is extremely challenging.
An indirect measurement has been proposed to constrain the mass and radius of NSs based on the pulse-profile modeling of the thermal emission from hot spots on the stellar surface of NSs~\citep{2016RvMP...88b1001W,2019ApJ...887L..26B,2012ARA&A..50..609W}. 
The characteristics of the observed pulse profile of a millisecond pulsar (MSP) are correlated to the mass, radius, and emission configurations of hot regions on the stellar surface via general relativity and special relativity effects, including the light bending, gravitational redshift, Doppler shifts, relativistic aberration, and propagation time differences.
Bayesian inference allows the estimation of the mass and radius of NSs conditional on the simultaneous spectral and timing measurements of soft x-rays.
It has been successfully applied to measure the mass and radius of two MSPs, PSR J0030+0451 and PSR J0740+6620, as demonstrated by the NICER collaboration \citep{2019ApJ...887L..21R, 2019ApJ...887L..24M, 2021ApJ...918L..28M, 2021ApJ...918L..27R}. 
Conversely, the constraints on mass and radius in x-ray flux oscillations in accreting pulsars are highly dependent on the adopted models~\citep{2012ARA&A..50..609W, 2015ApJ...808...31M,2016RvMP...88b1001W}. 
 
Despite the sensitivity of the NICER likelihood function to the geometry and temperature configurations of the surface hot regions, the background estimation is completely free when analyzing PSR J0030+0451~\cite{2019ApJ...887L..21R,2019ApJ...887L..24M}. 
It can cause under-estimation or over-estimation of the background when performing the NICER-only analysis, which consists of instrumental noise, cosmic background, and contamination sources around the target. 
To address this, Ref.~\cite{2021ApJ...918L..27R} conducted the joint analysis of the NICER and XMM-Newton datasets of PSR J0740+6620, where the phase-averaged x-ray data from XMM-Newton EPIC was utilized to constrain the unpulsed portion of the pulse profile.
Consequently, the XMM-Newton likelihood function significantly reduced the NICER posterior volume, impacting the inferred mass-radius and corresponding radiation geometries. 
The inclusion of XMM-Newton data in analyzing PSR J0740+6620 led to adjustments in the posteriors, favoring less compact stars. It highlighted the importance of background estimation through instrument cross-calibration.
The overall shift in the posterior probability distribution function (PDF) of the radius for 
PSR J0740+6620 was found to be much smaller than the measurement uncertainty, given the informative mass prior~\cite{2016ApJ...822...27M}.
This observed shift is anticipated to have an insignificant effect on EoS inference with typical EoS priors~\cite{2019MNRAS.485.5363G,2021ApJ...918L..29R}.

\begin{table*}
\centering
\caption{
Different x-ray measurements of masses and radii of NSs, along with the observations of the tidal deformability from the gravitational waves GW170817, used in the present Bayesian analysis. See Sec. \ref{sec:data} for details.}
\renewcommand\arraystretch{1.5}
\begin{ruledtabular}
\begin{tabular*}{\hsize}{@{}@{\extracolsep{\fill}}ccc@{}}
\multirow{2}{*}{PSR J0030+0451} & NICER only & $R=13.12_{-1.21}^{+1.35}\,{\rm km}$ and $M=1.41_{-0.19}^{+0.20}\,M_\odot$~\cite{2024ApJ...961...62V} \\
  & NICER x XMM-Newton & $R=14.44_{-1.05}^{+0.88}\,{\rm km}$ and $M=1.70_{-0.19}^{+0.18}\,M_\odot$~\cite{2024ApJ...961...62V} \\
\hline
  \multirow{2}{*}{PSR J0740+6620} & NICER only & $R=11.29_{-0.81}^{+1.20}\,{\rm km}$ and $M=2.078_{-0.063}^{+0.066}\,M_\odot$~\citep{2021ApJ...918L..27R} \\
  & NICER x XMM-Newton  & $R=12.39_{-0.98}^{+1.30}\,{\rm km}$ and $M=2.072_{-0.066}^{+0.067}\,M_\odot$~\citep{2021ApJ...918L..27R}\\
\hline
HESS J1731-347 & XMM-Newton x Suzaku & $R=10.4_{-0.78}^{+0.86}\,{\rm km}$ and $M=0.77_{-0.17}^{+0.20}\,M_\odot$~\cite{2022NatAs...6.1444D} \\
\hline
GW170817 & LIGO/Virgo & $\tilde\Lambda\leq 800$~\citep{2017PhRvL.119p1101A}
\end{tabular*}
\end{ruledtabular}
    \vspace{-0.4cm}
\label{table:detect precision}
\end{table*}

An interesting question arises regarding the effect of a joint analysis of the NICER and XMM-Newton data for PSR J0030+0451 on mass-radius inference.
In a recent study~\cite{2024ApJ...961...62V}, the radius and mass of PSR J0030+0451 were reanalyzed with an updated NICER response matrix and an upgraded analysis framework. 
To better estimate the background, i.e.\ the phase-invariant components that do not come from the hot regions on the stellar surface, the spectral data of the XMM-Newton EPIC MOS1 and MOS2 were considered. 
The inclusion of the XMM-Newton data is expected to yield a more robust constraint on the background, attributed to its imaging capability.
Substantial shifts in solution(s) were identified compared to those reported in Ref.~\cite{2019ApJ...887L..21R}.
Importantly, the newly introduced models infer configurations with one hot spot on the same hemisphere as the observer, a feature absent in Ref.~\cite{2019ApJ...887L..21R}.
The introduction of the XMM-Newton data in analyzing PSR J0030+0451 led to a reduction in background estimates from NICER-only data, resulting in higher compactness values~\cite{2021ApJ...918L..27R,2022ApJ...941..150S}.
In the absence of an informative mass prior from radio timing measurements of PSR 0030+0451, the overall shift of both mass and radius is evident and is anticipated to significantly impact the measurements, allowing for the constraining of the underlying NS EoS of NSs.

On the other hand, the challenge arises in reconciling the large radii of PSR J0740+6620 and PSR J0030+0451 with other observations, such as the tidal deformability inferred from the inspiring gravitational waves of GW170817~\cite{2018PhRvL.121p1101A,2019PhRvX...9a1001A}.
In particular, a recent analysis of the supernova remnant HESS J1731-347 suggests a very low mass and a small radius for the central compact object (CCO) within it~\citep{2022NatAs...6.1444D}.
Various theories and models have been proposed to interpret the observed mass and radius,  often incorporating exotic degrees of freedom beyond nucleons (see discussions in e.g., ~\citep{2024ApJ...966....3Y}), but none have yielded conclusive results thus far. 
Our interest extends to the exploration of multi-messenger observations and their potential to distinguish different phase states of the EoS modeling. 

In the present work, we are primarily interested in how these improved mass and radius constraints affect the EoS study of NSs with and without strangeness phase transitions\footnote{The present work will focus on the study of the EoS describing the NS core and does not discuss the properties of the NS crust.
In our analysis, the BPS~\citep{1971ApJ...170..299B} and NV~\citep{1973NuPhA.207..298N} EoSs are employed for the outer and inner crust, respectively.}.
The paper is organized as follows:
In Sec. \ref{sec:data}, we provide an overview of the relevant data utilized in this study. 
Sec. \ref{sec:model} offers a brief introduction to the fundamentals of EoS modeling. 
In Sec. \ref{sec:result} we present the results of our inference for the global NS properties and microscopic EoS parameters, for both nuclear matter and phase transitions. 
We conclude in Sec. \ref{sec:summary}.

\section{Multimessenger mass–radius constraints} 
\label{sec:data}

\subsection{Millisecond pulsars PSR J0030+0451 and PSR J0740+6620}

The mass-radius measurements of PSR J0030+0451 and PSR J0740+6620, as reported by the NICER collaboration, are considered in the present work. 
The contribution of the unpulsed component to the thermal emission of hot spots is constrained by jointly analyzing the NICER and XMM-Newton data.
Detailed various measurements are collected in Table \ref{table:detect precision}.

The Bayesian inference of the radius and mass of the rotation-powered MSP PSR J0030+0451 was initially performed independently by Miller et al.~\cite{2019ApJ...887L..24M} and Riley et al.~\cite{2019ApJ...887L..21R}, conditional on the pulse-profile data sets from the NICER observations. 
Different configurations of hot regions were proposed by Miller et al.~\cite{2019ApJ...887L..24M} and Riley et al.~\cite{2019ApJ...887L..21R}, respectively.
The two studies proposed different configurations of hot regions, with Miller et al.~\cite{2019ApJ...887L..24M} suggesting three oval, uniform-temperature emitting spots while Riley et al.~\cite{2019ApJ...887L..21R} proposed one small angular spot and the other being an azimuthally extended narrow crescent. 
Despite these differences in inferred hot region properties, the estimation of the equatorial radius and gravitational mass are consistent.
Miller et al.~\cite{2019ApJ...887L..24M} reported $R = 13.02_{-1.06}^{+1.24}\,\km$ and $M = 1.44_{-0.14}^{+0.15}\,\Msun$,
while Riley et al.\cite{2019ApJ...887L..21R} reported $R = 12.71_{-1.19}^{+1.14} \km$ and $M = 1.34_{-0.16}^{+0.15}\,\Msun$, at the 68\% confidence level.

As introduced earlier,  Ref.~\cite{2024ApJ...961...62V} recently employed an improved pipeline to conduct a joint analysis of the NICER and XMM-Newton datasets for PSR J0030+0451.
They employed four models to describe the surface heating distribution, with increasing complexity, \texttt{ST-U} (two single-temperature spherical caps), \texttt{ST+PST} (a single-temperature spherical cap and the other one protruding single-temperature), \texttt{ST+PDT} (a single-temperature spherical cap and the other one protruding dual-temperature) and \texttt{PDT-U} (two protruding dual-temperature spherical caps). 
The inferred radius and mass of the \texttt{ST-U} and \texttt{ST+PST} models are consistent with previous studies~\cite{2019ApJ...887L..24M,2019ApJ...887L..21R}, but are strongly disfavoured with smaller evidence values compared to those of \texttt{ST+PDT} and \texttt{PDT-U}.
The newly adopted models (\texttt{ST+PDT} and \texttt{PDT-U}) exhibit a higher preference in terms of Bayesian evidence. 
Conditioned on the most favored \texttt{PDT-U} model, NICER data alone yields $M=1.41_{-0.19}^{+0.20} \Msun$ and $R=13.12_{-1.21}^{+1.35}\km$. 
When incorporating data from both NICER and XMM-Newton, the derived values are $M=1.70_{-0.19}^{+0.18} \Msun$ and $R=14.44_{-1.05}^{+0.88}\km$. 
In this study, we employ a bivariate Gaussian distribution to emulate the two recent results:
\begin{equation}
    p(\boldsymbol{x}) = \frac{\exp\left[-\frac12(\boldsymbol{x}-\boldsymbol{\mu})^T\Sigma^{-1}(\boldsymbol{x}-\boldsymbol{\mu})\right]}{2\pi\sqrt{|\boldsymbol{\Sigma}|}}
\end{equation}
where $\boldsymbol{x}=(M,R)^T$. The mean vector $\boldsymbol{\mu}$ and the covariance matrix $\boldsymbol{\Sigma}$ are 
\begin{equation}
\boldsymbol{\mu} = \begin{pmatrix}
  \mu_M \\
  \mu_R \\
\end{pmatrix}, \quad
\boldsymbol{\Sigma} = \begin{pmatrix}
  \sigma_M^2 & \rho\sigma_M\sigma_R \\
  \rho\sigma_M\sigma_R & \sigma_R^2 \\
\end{pmatrix}
\end{equation}
For results using only NICER data, we use $\mu_M=1.41 \Msun$, $\mu_R=13.12\km$, $\sigma_M=0.20 \Msun$, $\sigma_R=1.3\km$ and $\rho=0.9$. While for results using both NICER and XMM-Newton data, we use $\mu_M=1.70 \Msun$, $\mu_R=14.44\km$, $\sigma_M=0.19 \Msun$, $\sigma_R=0.9\km$ and $\rho=0.9$. 

Similarly, we utilize the posteriors derived from the \texttt{ST-U} model for PSR J0740+6620, incorporating both NICER data alone and the joint analysis of NICER and XMM-Newton data.
In this case, the mass constraint of $2.08\pm0.07 \Msun$ obtained from radio timing measurement~\citep{2021ApJ...915L..12F} has been incorporated in the Bayesian inference,
leading to a more precise estimation of the mass-radius of NSs. 
When considering NICER data exclusively, the estimated equatorial radius and gravitational mass are are reported as $R=12.39_{-0.98}^{+1.30}\km$ and $M=2.072_{-0.066}^{+0.067} \Msun$ from~\citep{2021ApJ...918L..27R} and $R=13.71_{-1.50}^{+2.61}\km$ and $M=2.062_{-0.091}^{+0.090} \Msun$ from~\citep{2021ApJ...918L..28M}. 
With the inclusion of both NICER and XMM-Newton data, the inferred radius and mass are presented as $R=11.29_{-0.81}^{+1.20}\km$ and $M=2.078_{-0.063}^{+0.066} \Msun$ from~\citep{2021ApJ...918L..27R} and $R=11.51_{-1.13}^{+1.87}\km$ and $M=2.072_{-0.094}^{+0.087} \Msun$ from~\citep{2021ApJ...918L..28M}. 
For consistency, we adopt the results from~\citep{2021ApJ...918L..27R} in the present work. 

\subsection{Supernova remnant HESS J1731-347} \label{hess}
CCOs, located at the centres of supernova remnants, are radio-quiet isolated NSs emitting steady thermal x-ray. 
The majority of confirmed CCOs exhibit an absence of detectable pulsations~\citep{2023ApJ...944...36A},  indicative of a relatively uniform temperature distribution across the NS surface. 
Their energy spectra can be well-fitted with the blackbody model, even in the case of the three pulsed CCOs which exhibit non-uniform surface temperatures~\citep{2023ApJ...944...36A}.
These specific characteristics position CCOs as excellent 
sources for studying the EoS of dense matter~\citep{2022NatAs...6.1444D}. 
It is a challenge to generate such a simple thermal spectrum, considering the possible energetic physical processes occurring in the NS shell, atmosphere, or magnetosphere. 

The CCO in the supernova remnant HESS J1731-347 stands out as the brightest one in its class and attracts extensive observations. 
Analyzing XMM-Newton data, Ref.~\cite{2015A&A...573A..53K} found the blackbody spectrum model to be unfavorable, and hydrogen atmosphere models led to an unreasonably large distance. 
Instead, the carbon atmosphere model is compatible with the source location around 3\,kpc~\cite{2015A&A...573A..53K}.
A reanalysis by Ref.~\cite{2022NatAs...6.1444D} using XMM-Newton and Suzaku observations, with similar conditions as used in Klochkov et al.~\cite{2015A&A...573A..53K} (an energy range of $1-10\,\rm keV$, a uniform-temperature carbon atmosphere model, a distance of $3.2 \rm \,kpc$ and a wabs absorption model)
resulted in the same outcomes with slightly tighter parameter ranges. 
Including the low-energy band $<1\,\rm keV$ and substituting the outdated wabs absorption model with tbabs,
the NS mass was lowered from $\sim1.4 \Msun$ to $\sim1.0 \Msun$. 
However, the lack of a reliable distance introduces significant uncertainties in the NS parameters.
Ref.~\cite{2016MNRAS.458.2565D} provided evidence of an embedded optical star in the same shell as the CCO within the supernova remnant HESS J1731-347, suggesting that the optical star and the CCO are originally in a binary system. 
Estimating the CCO's distance based on that of the optical star, available from Gaia parallax measurements, yields $2.5\,\rm kpc$ (see \cite{2022NatAs...6.1444D} and references therein).
With this revised distance, Ref.~\cite{2022NatAs...6.1444D} further constrains the mass to $0.77^{+0.20}_{-0.17} M_{\odot}$ and the radius to $10.4^{+0.86}_{-0.78}\km$. When treating the distance as a free parameter with informative priors, the mass and radius parameters of the CCO are $M=0.83^{+0.17}_{-0.13} \Msun$ and $R=11.25^{+0.53}_{-0.37}\km$.

\subsection{GW170817}

The EoS governs not only the stable configuration of a single star but also the dynamics of NS mergers. During the inspiral phase, the influence of the EoS is evident in the tidal polarizability $\Lambda$, which can be computed by treating the tidal field from the companion as perturbations to general-relativistic hydrodynamic equation.
Apart from the mass and radius measurement, gravitational wave observations from binary NS mergers uniquely provide independent constraints on the EoS~\citep{2024ApJ...964...31M}.
In this study, we calculate the likelihood of gravitational wave event GW170817~\citep{2017PhRvL.119p1101A} using a high-precision interpolation method developed 
in~\citep{2020MNRAS.499.5972H}, which is derived from fitting the strain data released by LIGO/Virgo. 
Encapsulated in the {\sc PYTHON} package \textsf{toast}\footnote{https://git.ligo.org/francisco.hernandez/toast}, the likelihood function is expressed as:
\begin{equation}
    \mathcal{L}_{\rm GW} = F(M_{\rm ch},q,\Lambda_1,\Lambda_2)
\end{equation}
where $F(\cdot)$ represents the interpolation function.
$M_{\rm ch}$ and $q$ denote chirp mass and mass ratio respectively, while tidal deformability $\Lambda_i$ is evaluated from the component mass and the EoS, expressed as $\Lambda_i \equiv \Lambda_i(M_i;{\rm EoS})$. 
Note that $M_1 = M_{\rm ch}(1+q)^{1/5}/q^{3/5}$ and $M_2 = M_1q$.

\begin{figure}
    \vspace{-0.4cm}
\includegraphics[width=0.499\textwidth]{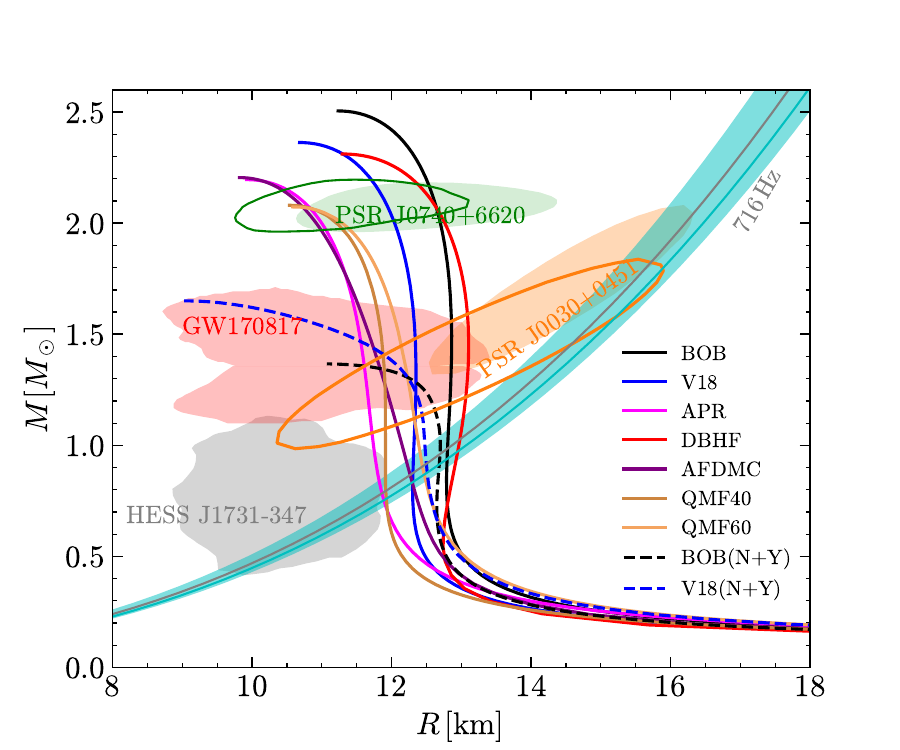}
     \vskip-2mm
\caption{Gravitational mass as a function of the stellar radius based on some representative EoS models with or without phase transition taken from Refs.~\citep{2021PrPNP.12003879B,2018ApJ...862...98Z} (see text for details). 
The mass-radius measurements from the NICER mission for the pulsars PSR J0030+0451 and PSR J0740+6620 are displayed, with the contours representing the results obtained solely from NICER data and the shaded areas reflecting the results obtained by utilizing both NICER and XMM-Newton data. 
Additionally, the mass-radius inferred from the GW170817 tidal deformability measurement and the mass-radius measurement of the CCO in the supernova remnant HESS J1731-347 are included.
All these measurements are presented at the 90\% confidence level. 
The grey line denotes the lower boundary of mass-radius, corresponding to the currently known maximum frequency $\nu =716\,{\rm Hz}$. The range for the minimum spin period of MSPs from Ref.~\citep{2024ApJ...962...80L} ($P_{\rm min}=1.43_{-0.09}^{+0.03}\,{\rm ms}$) is also presented at the 68\% confidence level. 
}
\label{fig:MR_of_diff_EOS}
\end{figure}

\subsection{Critical spin frequency} 
\begin{figure*}
    \vspace{-0.4cm}
\includegraphics[width=0.49\textwidth]{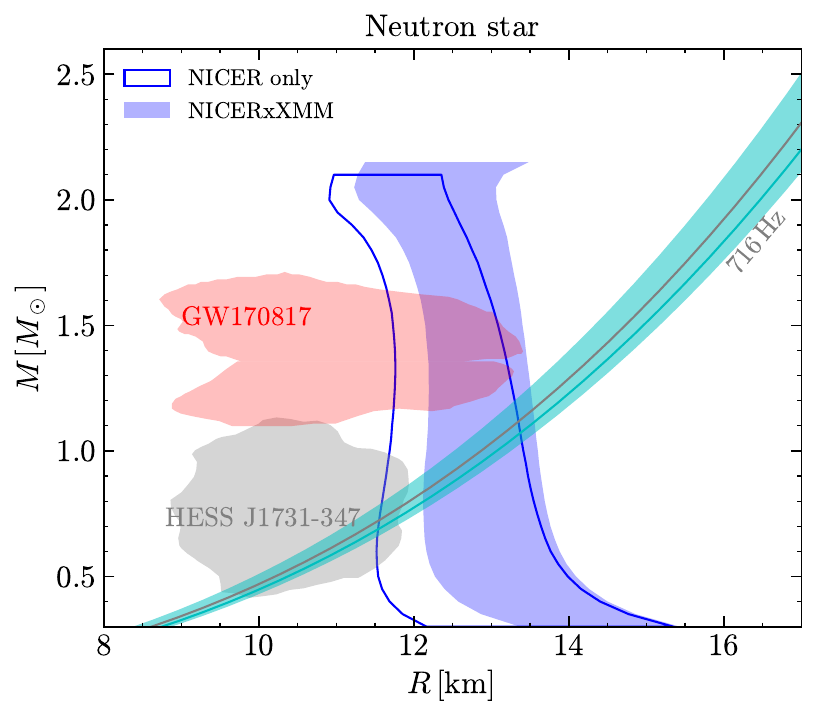}
\includegraphics[width=0.49\textwidth]{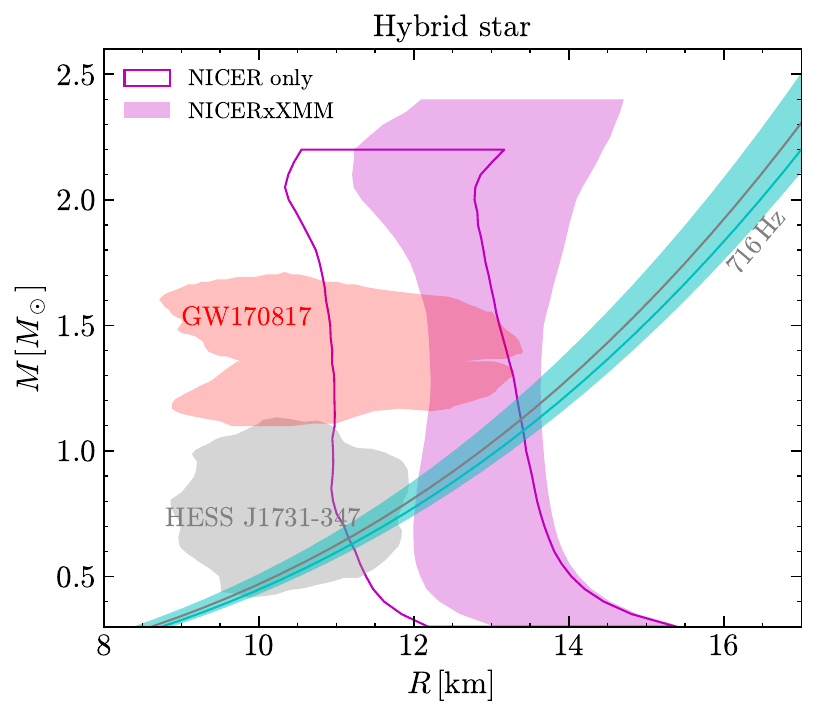}
\includegraphics[width=0.49\textwidth]{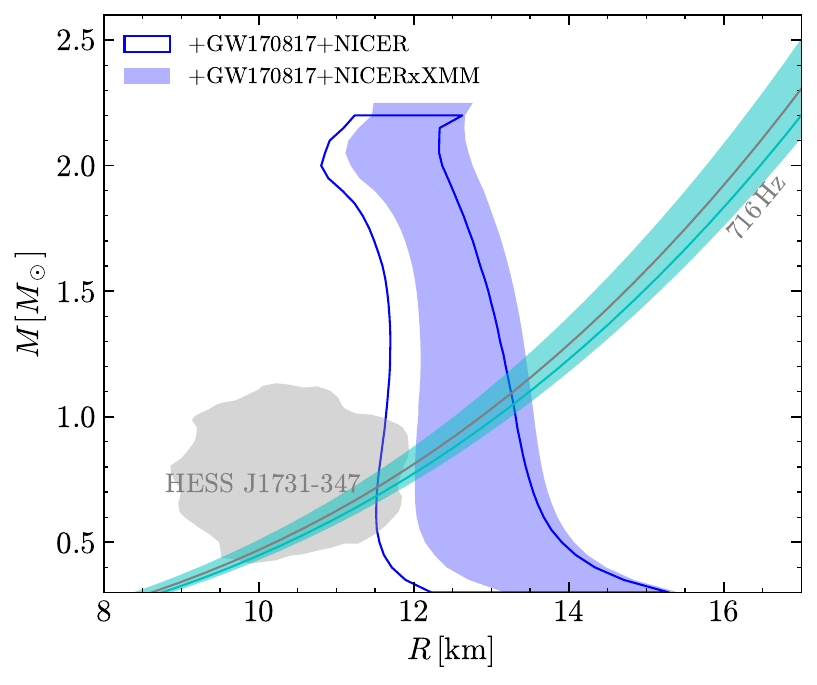}
\includegraphics[width=0.49\textwidth]{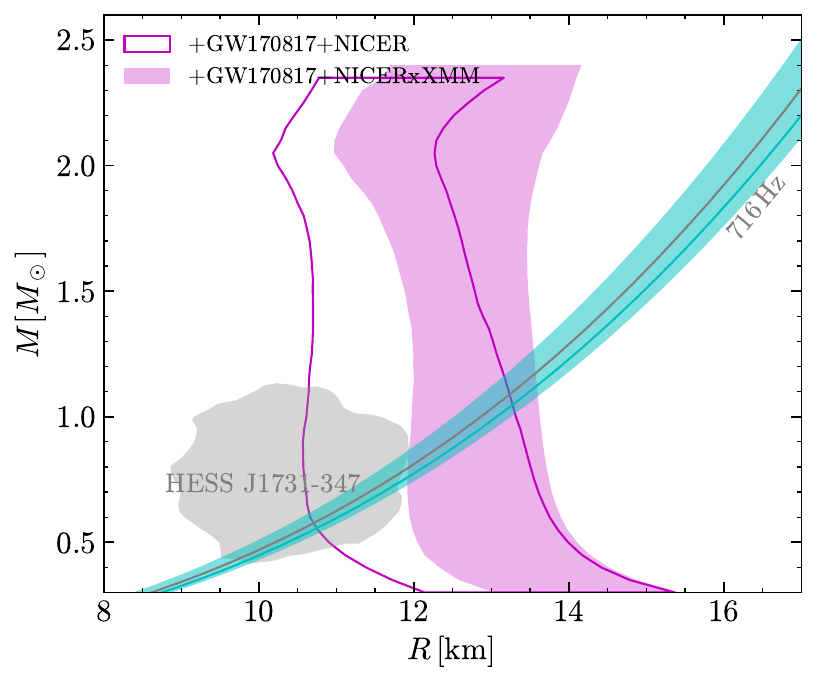}
     \vskip-2mm
\caption{Posteriors of mass-radius relations for NSs (left two panels) and HSs (right two panels), all at the 90\% confidence level. The upper two panels present the results without GW170817 constraints, while the lower two panels incorporate both NICER and GW170817 constraints. The results with XMM-Newton data as depicted within shaded regions show a noticeable overall shift towards the right when compared to those obtained without.
Other shaded regions correspond identically to those depicted in Fig.~\ref{fig:MR_of_diff_EOS}.
}
\label{fig:MR constraint}
\end{figure*}

The Keplerian frequency is one of the most studied physical quantities for pulsars~\citep{2021ApJ...910...62Z}. 
An EoS that predicts Kepler frequencies that are smaller than the observed rotational frequencies is to be rejected, as it is not compatible with observation.
The Kepler frequency, namely the maximum spin frequency, for an NS, is approximately given by $\nu_{\rm max} = 1045 \left(\frac{M}{M_\odot}\right)^{1/2} \left(\frac{R}{10{\rm km}}\right)^{-3/2}\,{\rm Hz}$, marking the threshold beyond which mass shedding initiates. 
This frequency depends significantly on the EoS governing the internal composition of pulsars~\citep{2009A&A...502..605H}.
The fastest known pulsar is PSR J1748−2446ad, with a frequency of $\nu=716\,{\rm Hz}$ (period $P=1.396$ ms). 
Recently, we studied the period distribution of MSPs~\citep{2024ApJ...962...80L} utilizing the largest available samples from known pulsar surveys and found no statistical support for a minimum spin period cutoff. 
Bayesian evidence favoring certain period distribution models suggests a minimum spin period of $P_{\rm min}= 1.43^{+0.03}_{−0.09} \,\rm ms$ is indicated.
This constraint serves as a lower limit in the mass-radius relationship (see Figs.~\ref{fig:MR_of_diff_EOS} and \ref{fig:MR constraint}),
with no capacity yet to distinguish between different models of EoSs.

\begin{figure*}
\includegraphics[width=0.49\textwidth]{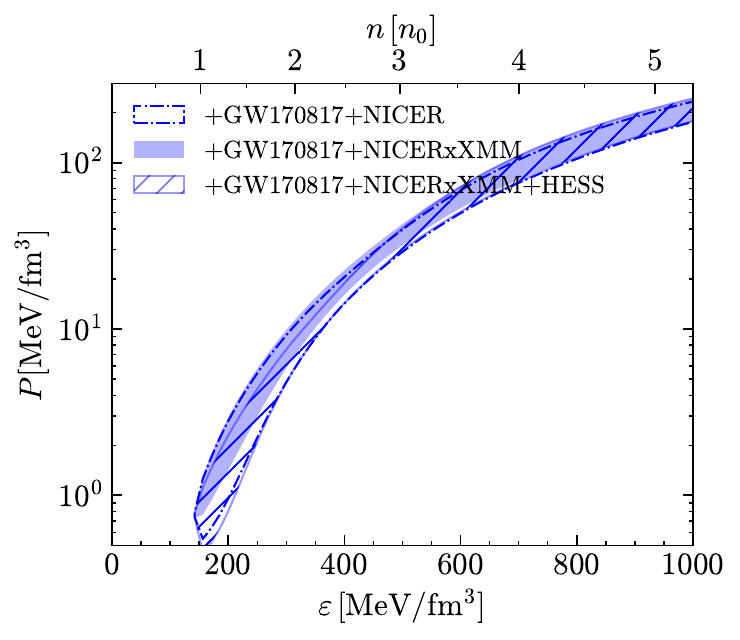}
\includegraphics[width=0.49\textwidth]{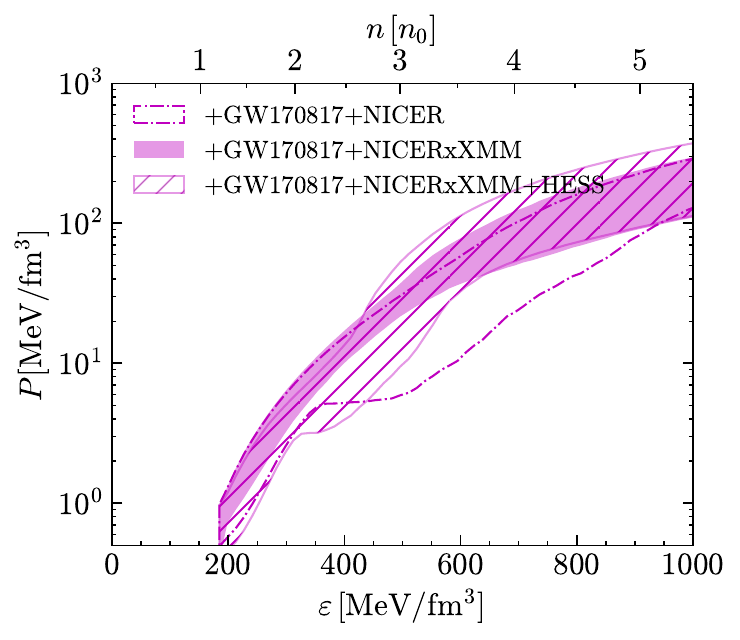}
\includegraphics[width=0.49\textwidth]{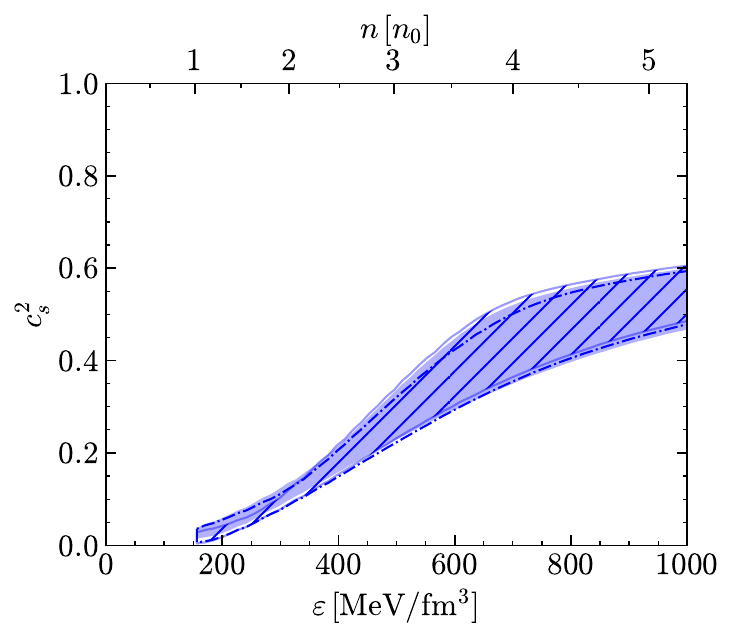}
\includegraphics[width=0.49\textwidth]{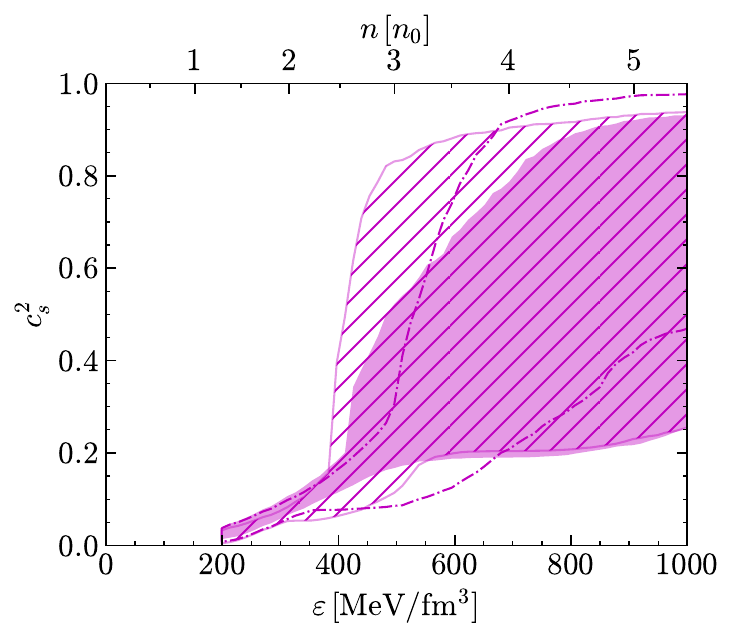}
     \vskip-2mm
\caption{Posteriors of EoSs and the sound speed squared in the matter of NSs (left two panels) and HSs (right two panels), all at the 90\% confidence level. 
The additional upper horizontal axis in each panel indicates the corresponding baryon number density within the most favored EoS parameter sets considering both GW170817 and NICER with XMM-Newton data..
}
\label{fig:EOS constraint}
\end{figure*}

\section{Theoretical modeling of EoS}
\label{sec:model}

Before delving into our inferences on the EoS, we observe the key constraints detailed above in Sec.~\ref{sec:data}, which are illustrated in Fig.~\ref{fig:MR_of_diff_EOS}.
These constraints are juxtaposed with several representative microscopic EoS models (as taken from the review article~\citep{2021PrPNP.12003879B}), starting from realistic nucleon-nucleon potentials (Argonne V18~\citep{1995PhRvC..51...38W}, Bonn B~\citep{1990PhRvC..42.1965B}) explicitly including many-body forces; BOB (N+Y) and V18 (N+Y) are EoSs containing hyperons.
Two phenomenological relativistic-field EoSs are QMF40 and QMF60 from Ref.~\cite{2018ApJ...862...98Z}, where the isospin-asymmetric contribution of the nuclear EoS, characterized by the symmetry energy $E_{\rm sym}(n)$, is purposely changed to be studied later:
\begin{eqnarray}
E/A (n, \beta) =  E/A (n, \beta=0) + E_{\rm sym}(n)\beta^2 + ...
\label{eq:snm}
\end{eqnarray}
where $n=n_n+n_p$ is the total nucleon number density and $\beta = (n_n-n_p)/n$ is the isospin asymmetry of nuclear matter. $E/A = \varepsilon/n- m_N$ is the energy per nucleon, $m_N=939\MeV$ being the nucleon mass in free space.

The minimum of the $E/A(n)$ curve for symmetric nuclear matter ($\beta = 0$) corresponds to a bound equilibrium state at zero pressure. 
The values of $E/A$ and $n$ at this minimum will be denoted by $E_0/A$ and $n_0$, where $n_0$ is the so-called saturation density $n_0=0.16\fm3$ resulting from the interplay of the short-distance repulsion and the long-distance attraction in the nucleon-nucleon interaction.
Around the saturation density $n_0$, both $E/A (n, \beta=0)$ and $E_{\rm sym}(n)$ can be expanded in isospin asymmetry $\beta$ as follows:
\begin{eqnarray}
&&E/A (n,0) = E_0/A +\frac{1}{18} K_0\frac{n-n_0}{n_0}  + ...\\
&&E_{\rm sym}(n) = E_{\rm sym}^0 + \frac{1}{3} L_0\frac{n-n_0}{n_0} + ... 
\end{eqnarray}
where $K_0$, $E_{\rm sym}^0$ and $L_0 = 3n_0 ({dE_{\rm sym}}/{dn})_{n_0}$ are the incompressibility, the symmetry energy and its slope at the saturation point, respectively.
$K_0$ gives the curvature of $E_0/A$ at $n = n_0$ and the associated increase of the energy per nucleon of symmetric nuclear matter due to a small compression or rarefaction, whereas $E_{\rm sym}^0$ determines the increase in the energy per nucleon due to a small asymmetry $\beta$ in asymmetric nuclear matter.

\begin{figure*}
    \vspace{-0.4cm}
\includegraphics[width=0.49\textwidth]{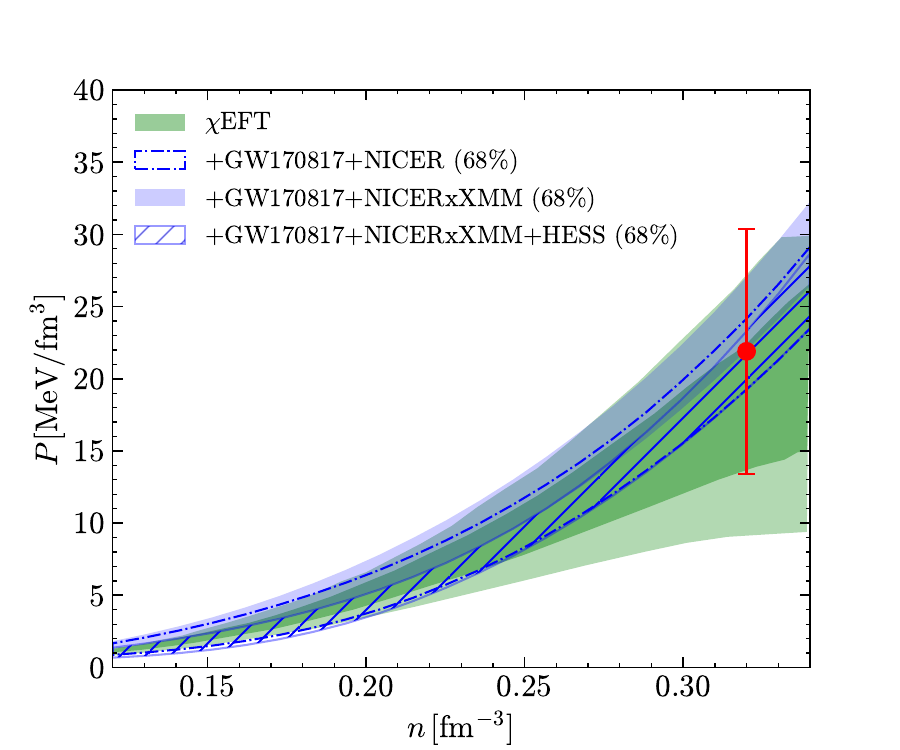}
\includegraphics[width=0.49\textwidth]{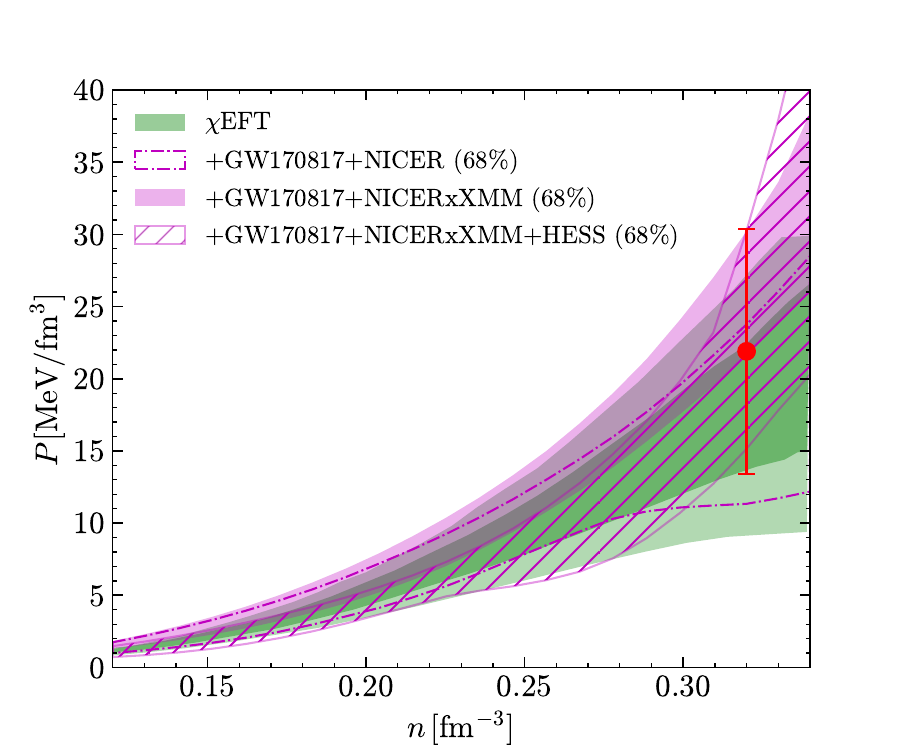}
     \vskip-2mm
\caption{Posteriors of baryon number density vs. pressure for NSs (left panel) and HSs (right panel). Also shown are the 68\% and 95\% credible regions derived in Ref.~\citep{2021PhRvC.103d5808D} using N$^3$LO $\chi$EFT calculations~\citep{2020PhRvL.125t2702D,2020PhRvC.102e4315D}. The data points in both panels with error bars depict the inferred pressure at $\sim2\,n_0$ from Ref.~\citep{2021PhRvD.104f3003L} without including the updated radius measurement of J0030+0451 incorporating the XMM-Newton data.
}
\label{fig:P-n}
\end{figure*}

Strong interactions play a dominating role and are the basic ingredient of calculations in the NS EoS, i.e., $P(\varepsilon)=n(d\varepsilon/dn) - \varepsilon$.
The theoretical underpinning of strong interactions between baryons is the quantum chromodynamics (QCD), where the fundamental fields are those of quarks and gluons. 
A comprehensive exploration of constructing baryonic interactions based on lattice QCD is available in reviews such as \citep{2011PrPNP..66....1B,2012PTEP.2012aA105A}. However, lattice QCD calculations are currently extremely expensive and often limited to scenarios with large quark masses.
An alternative approach, rooted in the realm of quark degrees of freedom, is formulated within the framework of effective field theory~\citep{1990PhLB..251..288W,1991NuPhB.363....3W}, developed as chiral effective field theory ($\chi$EFT) approach~\citep{2011PhRvC..83c1301H}. This approach organizes terms based on their dependence on the physical parameter $q/m_N$, where $m_N$ denotes the nucleon mass and $q$ represents a generic momentum in the relevant Feynman diagrams. This systematic ordering facilitates the calculation of various terms constituting the baryon force, yielding nucleon-nucleon interactions that reasonably reproduce two-body data. Extending chiral perturbation theory to nuclear matter calculations has been a subject of considerable effort.
However, the assumption of a small $q/m_N$ parameter inherently limits the applicability of these forces, with the safe maximum density typically around the saturation value, $n_0$. Extrapolation becomes essential for densities beyond $n_0$, and theoretical uncertainty grows rapidly. Several calculations have extended up to $2\,n_0$ (e.g., \citep{2020PhRvL.125t2702D,2020PhRvC.102e4315D}), with uncertainties estimated by analyzing the order-by-order convergence in the chiral expansion and many-body perturbation theory. 
See more discussions in e.g.,~\citep{2021PrPNP.12003879B,LInpr}.
In the ensuing sections, we provide illustrative examples of their application to the study of NS EoS, as depicted in Fig.~\ref{fig:P-n}.

Currently, it remains a formidable challenge to connect QCD with the low-energy nuclear physics phenomena since it is in the non-perturbative regime. Consequently, one has to resort to phenomenological approaches.
Starting in the 1950s, the development of meson-exchange models for nucleon interaction marked a significant step forward, grounded in Yukawa's idea.
Namely, the strong interaction among the different baryons originates from the coupling of baryons to various meson fields, where quarks and gluons do not appear explicitly.
Correspondingly, such meson-exchange interaction models operate with the baryon and meson fields, and the meson couplings are described by corresponding Lagrangian densities, depending on the symmetry behavior of a meson field under rotations and reflections. 
The inclusion of mesons with rest mass below $1 \,\rm GeV$ suffices for these models.
In the meson-exchange picture, one can construct the bare/realistic interaction to describe nucleon-nucleon scattering data and phase shifts at laboratory energies $\leq 350 \MeV$, like the above-mentioned Argonne V18, Bonn B interactions, serving as fundamental inputs for microscopic many-body frameworks;
One can also construct in-medium/effective interactions, wherein phenomenological parameters reproduce various properties of nuclear systems like finite-nuclei and NS observations.
By employing the latter effective interactions, one can compute the energy of hadronic systems through mean-field approximation. It is the framework of relativistic mean-field model~\citep{1974AnPhy..83..491W,1992RPPh...55.1855S} that is widely adopted in NS physics.
Its Lorentz invariance guarantees that the sound speed in dense matter does not exceed the speed of light at any density. 
The mean-field approximation, applied to a spatially uniform ground state of hadronic matter, entails hadrons occupying momentum states within Fermi spheres. The eigenfunctions of hadrons manifest as plane-wave solutions to the Dirac equations.
The self-consistent solution involves solving the Klein-Gordon equations for meson fields alongside the Dirac equations for the hadrons. The many-body state is subsequently constructed as an independent particle, derived from the single-particle wave functions. Hadrons are represented by four-component Dirac spinors, and meson-field operators are substituted with their expectation values in the ground state. 

In the computation of the NS EoS, the selected theoretical approaches also depend on the relevant degrees of freedom of the problem, from nuclei and nucleons at lower densities to additional particles, such as hyperons and quarks, at high densities~\citep{2008IJMPE..17.1635L,2016PhRvC..94d5803Z}.
In the simplest and conservative picture, the NS core is modeled as an electrically neutral uniform fluid of neutrons, protons, electrons and muons in equilibrium with respect to the weak interaction ($\beta$-stable nuclear matter).
The isoscalar scalar $\sigma$ meson and the isoscalar vector $\omega$ meson mediate the long and short-range part of the interaction, respectively, in symmetric nuclear matter, while isovector
mesons (like the isovector vector $\rho$ meson) need to be included as well to treat isospin-asymmetric matter.
One Lagrangian density based on such effective interaction can be written as:
\begin{eqnarray}
\mathcal{L}& = & \overline{\psi}\{(i\gamma_\mu \partial^\mu - m^\ast - \gamma_{\mu} (g_{\omega_N}\omega^{\mu} + \frac{1}{2}g_{\rho_N}\vec{\rho}_{\mu}\vec{\tau})\}\psi  \nonumber \\
           && +\frac{1}{2}\partial^{\mu}\sigma\partial_{\mu}\sigma- \frac{1}{2}m_\sigma^2 \sigma^2 - \frac{1}{3} g_2\sigma^3 - \frac{1}{4}g_3\sigma^4 \nonumber \\
           && - \frac{1}{4}W^{\mu\nu}W_{\mu\nu}+ \frac{1}{2}m_\omega^2\omega^{\mu}\omega_{\mu} \nonumber \\
           & & - \frac{1}{4}\vec{R}^{\mu\nu}\vec{R}_{\mu\nu}
           + \frac{1}{2}m_{\rho}^2\vec{\rho}^{\mu}\vec{\rho}_{\mu} \nonumber \\
           && + \Lambda_v (g_{\omega N}^2\omega^{\mu}\omega_{\mu}) (g_{\rho N}^2\vec{\rho}^{\mu}\vec{\rho}_{\mu})
           \ ,
           \label{eq:L}
\end{eqnarray}
where $\psi$ represents the wave function of nucleons. $\sigma$, $\omega_{\mu}$, $\vec{\rho}_{\mu}$ denote the fields of $\sigma$, $\omega$, and $\rho$ mesons, respectively. $W^{\mu\nu}$ and $\vec{R}^{\mu\nu}$ are the anti-symmetry tensor fields of $\omega$ and $\rho$ mesons.
Non-linear meson self-interaction terms in the Lagrangian are introduced to account for medium dependence of the effective mean-field interactions (see e.g.,~\citep{1977NuPhA.292..413B}).
The cross-coupling from the $\omega$ meson and $\rho$ meson is introduced to achieve a reasonable slope of symmetry energy~\citep{2001PhRvL..86.5647H}.
$m_{\sigma} = 510~\rm{MeV}$,~$m_{\omega}=783~\rm{MeV}$, and $m_{\rho}=770~\rm{MeV}$ are the meson masses.
$g_{\omega N}$ and $g_{\rho N}$ are the nucleon coupling constants for $\omega$ and $\rho$ mesons. 
From the simple quark counting rule, we obtain $g_{\omega N}=3g_{\omega q}$ and $g_{\rho N}=g_{\rho q}$
from the corresponding coupling constants with quarks. 
 $m^\ast=m_N-g_{\sigma_N} \sigma$ is the (Dirac) effective nucleon mass (see discussions in e.g., \citep{2016PhRvC..93a5803L}). 
Simply, $m^\ast$ can be evaluated from the calculation of the confined quarks as a function of the $\sigma$ field, $g_{\sigma N} = -\partial m^\ast/\partial \sigma$, which defines the $\sigma$ coupling with nucleons (depending on the parameter $g_{\sigma q}$).
For example, by constructing a nucleon from confined quarks with a two-body confining potential in a quark mean-field (QMF) model~\citep{2018PhRvC..97c5805Z}, the mass of the nucleon in the nuclear medium is expressed as the binding energy of three quarks $E_0=\sum_q\epsilon_q^\ast$, defined by the zeroth-order term after solving the Dirac equation of each quark inside a nucleon in a (harmonic oscillator) confinement potential~\cite{2018ApJ...862...98Z}:
\begin{eqnarray}
m^\ast=E_{0}-\epsilon_{\rm c.m.}+\delta m^\pi+(\Delta E)_g \ ,
\label{eq:m*}
\end{eqnarray}
where the effective single quark energy is given by $\epsilon_q^*=\epsilon_{q}-g_{q\omega}\omega-\tau_{3q}g_{q\rho}\rho$, and the effective quark mass is given by $m_q^\ast = m_q-g_{\sigma q}\sigma$.
In Eq.~(\ref{eq:m*}), we shall consider three corrections in the zeroth-order nucleon mass in the nuclear medium, including the contribution of the center-of-mass (c.m.) correction $\epsilon_{\rm c.m.}$, pionic correction $\delta m^\pi$ and gluonic correction $(\Delta E)_g$.
The pion correction is generated by the chiral symmetry of QCD theory and the gluon correction by the short-range exchange interaction of quarks. See details in~\citep{2020JHEAp..28...19L,2023PhRvC.108b5809Z}.

\begin{figure*}
    \vspace{-0.4cm}
\includegraphics[width=0.99\textwidth]{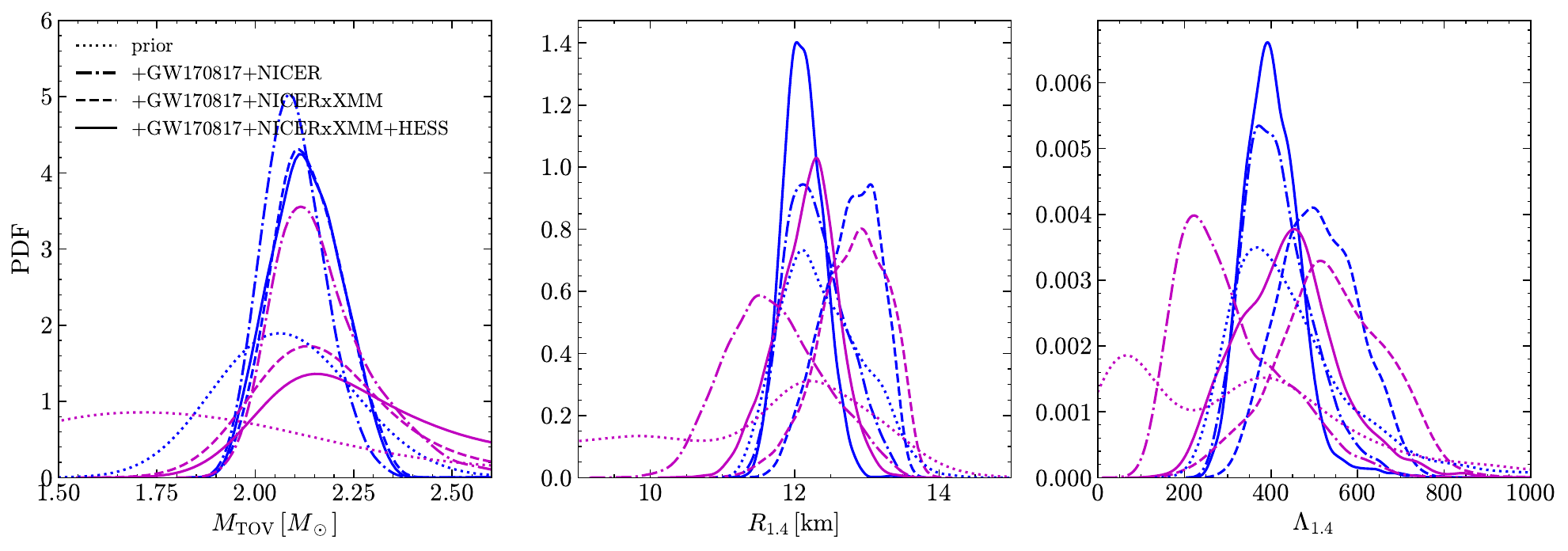}
     \vskip-2mm
\caption{Posterior distributions of the stellar properties: the maximum mass $M_{\rm TOV}$, the radius for a $1.4\Msun$ star $R_{1.4}$ and the tidal deformability of a $1.4\Msun$ star $\Lambda_{1.4}$, conditioned by different constraints as outlined in Sec.~\ref{sec:data}. The priors are also shown in dotted curves. 
}
\label{fig:star params pdf}
\end{figure*}

Following standard quantum field theory, with the Euler-Lagrange equation, the equations of motion of the nucleon and mesons can be obtained and subsequently solved self-consistently in terms of the mean-field approximation.
Note that the space components of the vector mesons are removed in the parity conservation system.
In addition, the spatial derivatives of nucleons and mesons are neglected in the infinite nuclear matter due to the transformation invariance.
The energy density and pressure are then generated by the energy-momentum tensor $T_{\mu\nu}$ and are then calculated as their ground-state expectation values,
\begin{eqnarray}
T_{\mu\nu} = -g_{\mu\nu} \mathcal{L} + \frac{\partial\phi_i}{\partial x^{\nu}} \frac{\mathcal{L}}{\partial(\partial\phi_i)/\partial x_{\mu}} \ ,
           \label{eq:tensor}
\end{eqnarray}
where $\phi_i$ denotes the nucleon and various mesons.

The relativistic-field EoS model detailed above is characterized by six parameters ($g_{\sigma q}, g_{\omega q}, g_{\rho q}, g_2, g_3, \Lambda_v$).
To determine these parameters, we reproduce the six experimental parameters of nuclear matter: 
the saturation density $n_0$ and the corresponding values at the saturation point of the binding energy $E_0/A$, the incompressibility $K_0$, the symmetry energy $E_{\rm sym}^0$, the symmetry energy slope $L_0$ and the nucleon effective mass $m^\ast_0$.
In the subsequent section, following a methodology akin to our prior work in~\citep {2023PhRvC.108b5809Z,2023ApJ...943..163Z}, we vary four parameters ($K_0, E_{\rm sym}^0, L_0, m^\ast_0$), while keeping the saturation density $n_0$ and energy per nucleon $E_0/A$ are constants. 
Uniform prior distributions within the empirical ranges, as presented in Table~\ref{tb:comparison2}, are employed for these parameters.
For instance, for the symmetry energy slope $L_0$, besides the two typical values of 40 and 60 MeV (see discussions in~\citep{2023Parti...6...30L}) as illustrated in Fig.~\ref{fig:MR_of_diff_EOS}, a broad range from 20 to 120 MeV is considered to respect the somewhat contradictory results from PREX-II experiment~\citep{2021PhRvL.126q2502A} and $\chi$EFT calculation~\citep{2022NatPh..18.1196H}.

As for NSs with a transition to deconfined quark phase in their interior (at some transition pressure $P_{\rm t}$), same with our previous works~\cite{2020ApJ...904..103M,2021ApJ...913...27L}, we employ the general constant-speed-of-sound parametrization~\cite{2013PhRvD..88h3013A} from the transition onset up to the maximum central pressure of a star (see detailed discussions in e.g., \citep{2021ChPhC..45e5104X}), while keeping using QMF for the hadronic matter:
\begin{equation}
\ep(P) = \left\{\!
\begin{array}{ll}
\ep_{\rm t}(P) & P<P_{\rm t} \\
\ep_{\rm t}(P_{\rm t})+\Delta\ep+c_{\rm s}^{-2} (P-P_{\rm t}) & P>P_{\rm t}
\end{array}
\right.\
 \nonumber \\
\label{eq:CSS_EoS}
\end{equation}
In addition to the four EoS parameters used in the QMF model, we introduce three additional parameters for the quark matter phase, i.e., the transition density $n_{\rm t}/n_0$, the transition strength $\Delta\varepsilon/\varepsilon_{\rm t}$ and the sound speed squared in quark matter $c_{\rm s}^2$. 
The priors for these parameters are chosen based on Ref.~\citep{2021ApJ...913...27L} and are presented in Table~\ref{tb:comparison2}. 
Note that the value of the sound speed can be varied from $1/\sqrt{3}$ (the perturbative QCD value) to 1 (the casual limit).

\begin{table}[t]
  \caption {Most probable intervals of the stellar properties (quoted as median+68\% credible interval): the maximum mass $M_{\rm TOV}$ in $\Msun$, the radius of a $1.4\Msun$ star $R_{1.4}$ in $\km$ and the dimensionless tidal deformability of a $1.4\Msun$ star $\Lambda_{1.4}$.
  }
    \setlength{\tabcolsep}{0.8pt}
\renewcommand\arraystretch{1.3}
\begin{tabular*}{\hsize}{@{}@{\extracolsep{\fill}}lccc@{}}
    \hline     \hline
     &$R_{1.4}$ & $\Lambda_{1.4}$ & $M_{\rm TOV}$\\
     \hline
         NS (w.o. XMM) &$12.24_{-0.36}^{+0.49}$ &$406.10_{-64.56}^{+88.31}$ &$2.09_{-0.07}^{+0.08}$ \\ 
        NS (with XMM) &$12.82_{-0.46}^{+0.36}$ &$506.88_{-88.41}^{+97.11}$ &$2.13_{-0.08}^{+0.10}$ \\   
        NS (plus HESS) &$12.11_{-0.26}^{+0.29}$ &$398.44_{-57.14}^{+63.66}$ &$2.13_{-0.09}^{+0.10}$\\            
    \hline
    HS (w.o. XMM)  &$11.69_{-0.63}^{+0.75}$ &$277.69_{-86.50}^{+161.95}$ 
    &$2.15_{-0.09}^{+0.16}$ \\ 
    HS (with XMM)  &$12.85_{-0.51}^{+0.45}$ &$534.71_{-120.78}^{+136.63}$ &$2.21_{-0.14}^{+1.15}$ \\ 
    HS (plus HESS) & $12.20_{-0.47}^{+0.35}$ &$437.62_{-118.18}^{+104.86}$ &$2.35_{-0.25}^{+0.94}$  \\         
    \hline     \hline
\end{tabular*}
          \vspace{-0.2cm}
  \label{tb:comparison1}
\end{table}

From Fig.~\ref{fig:MR_of_diff_EOS}, it is evident that the radius of a $1.4\Msun$ star, denoted as $R_{1.4}$, is subject to nuclear matter properties, specifically $L_0$ at the saturation point; 
The EoS models QMF40 and QMF60, discussed previously, belong to a set of NS EoSs constructed in Ref.~\cite{2018ApJ...862...98Z}, enabling fine-tuning of the symmetry energy slope $L_0$ and a systematic exploration of the $R$ (or $\Lambda)$ versus $L$ relationship. 
with well-reproduced robust observables from laboratory nucleons, nuclear saturation, heavy-ion collisions (HIC), and heavy pulsars.
Rather, the maximum mass and the possibility of a strangeness phase transition (e.g., involving hyperons or deconfined quarks) are associated with the EoS at supersaturation densities.
The XMM-Newton data of PSR J0030+0451, in conjunction with the NICER data, can shed light on the EoS at low density (manifested by $L_0$) at $\sim n_0$, intermediate density (manifested by $R_{\rm 1.4}$) at $\sim 2\,n_0$ and high density (manifested by the maximum mass $M_{\rm TOV}$) at $\sim 5\,n_0$. 
On the other hand, the mass and radius data of HESS J1731-347 should primarily complement other EoS constraints at low and intermediate densities at $\sim 1$-$2\,n_0$.
It is expected that the stiffening effect introduced by the XMM-Newton data and the softening effect introduced by the HESS J1731-347 data will lead to a compromise, resulting in a flattening of the credible ranges for both global NS properties and microscopic EoS parameters.

\section{Bayesian inference of EoS}
\label{sec:result}

\begin{figure*}
    \vspace{-0.2cm}
\includegraphics[width=0.99\textwidth]{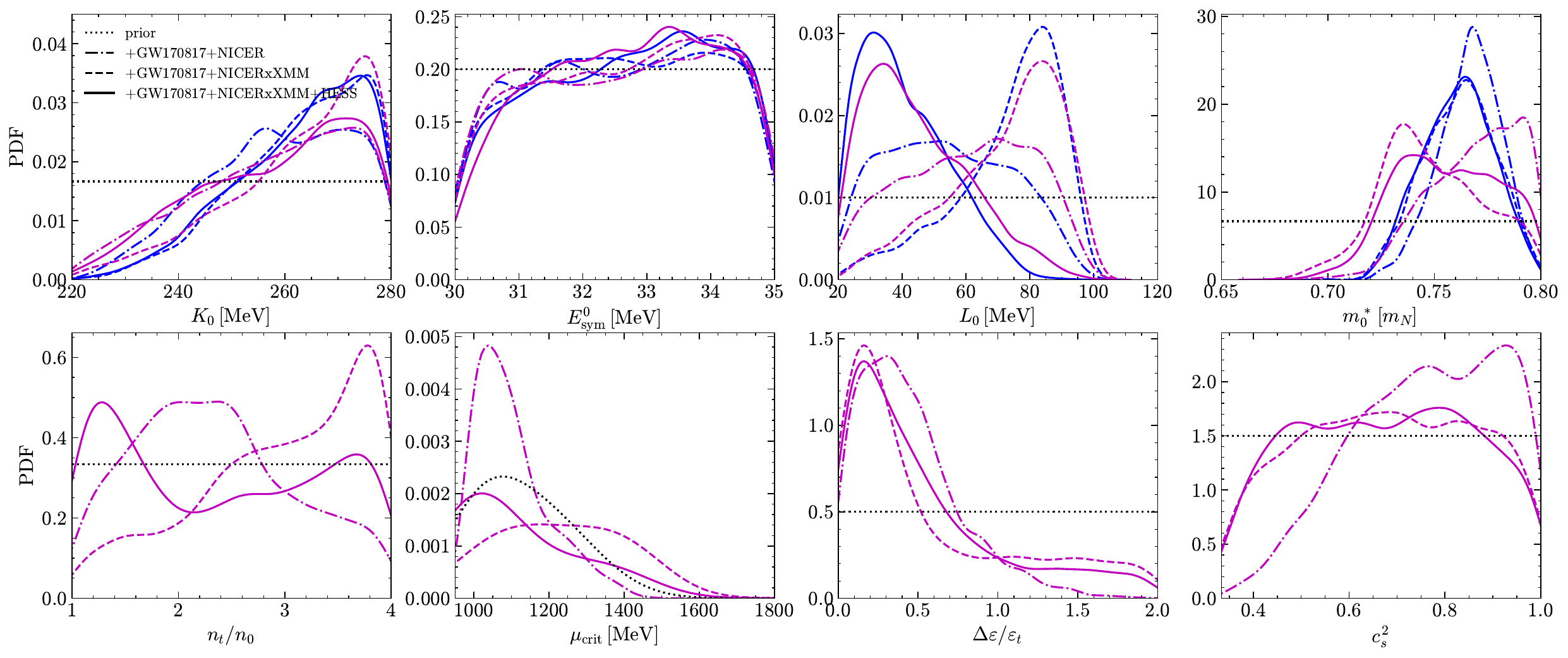}
     \vskip-2mm
\caption{Same as Fig.~\ref{fig:star params pdf}, but for various properties for nuclear matter saturation (the incompressibility $K_0$ (in MeV), the symmetry energy $E_{\rm sym}^0$ (in MeV), the symmetry energy slope $L_0$ (in MeV) and the ratio between the effective mass and free nucleon mass $m^\ast_0/m_N$) and phase transition (the critical density ratio $n_{\rm t}/n_0$, the critical chemical potential $\mu_{\rm crit}$ (in MeV), the transition strength $\Delta\varepsilon/\varepsilon_{\rm t}$, and the squared sound velocity $c_s^2$) in high-density (quark) matter. See text in Sec.~\ref{sec:model}.
}
\label{fig:EOS params pdf}
\end{figure*}

In this section, we present the outcomes of our Bayesian inference regarding NS and EoS properties. Our analysis incorporates astronomical observations within the Bayesian framework detailed in~\citep{2023PhRvC.108b5809Z} and \citep{2021ApJ...913...27L}. 
We report stellar properties in Figs.~\ref{fig:MR constraint}, \ref{fig:star params pdf} and Table~\ref{tb:comparison1}. Various EoS properties are detailed in Figs.~\ref{fig:EOS constraint}, \ref{fig:P-n}, \ref{fig:EOS params pdf}, and Table~\ref{tb:comparison2}.
We discuss primarily the impact of the updated mass and radius analysis for PSR J0030+0451 data and explore as well how different measurements of NSs with low and high masses jointly narrow the EoS parameter space.
The current quantitative results (at some confidence level) of the EoS at supersaturation density are also confronted with the extrapolation of the microscopic calculations of the EoS e.g., in the $\chi$EFT method, and general improvement is achieved (see below in Fig.~\ref{fig:P-n}).

Analyzing Fig.~\ref{fig:MR constraint}, it becomes apparent that a tension exists for $R_{1.4}$ between the constraints derived from the NICER sources and HESS J1731-347. This tension appears more pronounced than the one observed between the NICER sources and GW17817, a notable point frequently discussed in existing literature. The incorporation of XMM-Newton data for the masses and radii of two NICER sources further intensifies this tension.
As a consequence, we anticipate that the uncertain ranges of $R_{1.4}$ (as well as $L_0$) will experience effective narrowing through the combination of multimessenger data, incorporating radio, gravitational wave, and x-ray observations.

In Fig.~\ref{fig:EOS constraint}, we present the EoS and sound speed for both NSs (left panels) and HSs (right panels). The updated radius measurements, incorporating XMM-Newton data, enhance the precision of the EoS inference by tightening the constraint on the pressure at densities around $\sim1$-$2\,n_0$ ($\sim1$-$5\,n_0$) for NSs (HSs).
EoS parameter spaces are much enlarged when considering possible strangeness phase transitions and large uncertainties reside especially at high densities above $\sim2\,n_0$, leading to unresolved predictions for the maximum mass ($M_{\rm TOV}$) and the sound speed at high-density cores of NSs.
For instance, the inclusion of XMM-Newton data for the masses and radii of two NICER sources raises $M_{\rm TOV}$ from $2.09\Msun$ ($2.15\Msun$) to $2.13\Msun$ ($2.21\Msun$) in the case of pure NS (HS), as depicted in Fig.~\ref{fig:MR constraint}. 
If we further incorporate HESS data, we observe an enlargement of the EoS parameter space for both NSs and HSs due to the compromising effects mentioned earlier.
In Fig.~\ref{fig:star params pdf}, the compromising effects introduced by the XMM data (comparing dash-dotted curves with dashed ones) and the HESS data (comparing dashed curves with solid ones) become more evident. Consequently, $R_{1.4}$ is refined to a much narrower range around $12.11\km$ ($12.20\km$) for pure NSs (HSs). A similar trend is observed for $\Lambda_{1.4}$. Nevertheless, the posterior PDF of $M_{\rm TOV}$ exhibits flattening toward higher values. For instance, the $1\sigma$ upper bound increases from $2.31\Msun$ with only the NICER data to over $\sim3.3\Msun$ when adding the XMM and/or HESS data (see Table \ref{tb:comparison1}).

From Fig.~\ref{fig:P-n}, it is seen that the updated mass and radius measurements, incorporating XMM-Newton data, do not inform the EoS within the applicable regime of $\chi\text{EFT}$ (i.e., below $n_0$), and effectively lift the lower bounds of the pressure above $n_0$ in both cases of pure NSs and HSs.
Consequently, by combining the information of $\chi\text{EFT}$ at subsaturation densities with multiple observations of radio, gravitational wave, and x-ray data at high densities, the constraint on the pressure at densities $\sim1$-$3\,n_0$ is effectively tightened.
The posterior +GW170817+NICER (\texttt{w.o.XMM})  is comparable to the seventh column of Table II in Ref.~\citep{2021PhRvD.104f3003L}, where the mass and radius constraints for PSR J0030+0451 and J0740+6620 from NICER are utilized, but without the updated radius measurement of J0030+0451 incorporating the XMM-Newton data. 
The current analysis with \texttt{NICER$\times$XMN} for light PSR J0030+0451 sets a much narrower limit on the stiffness at $\sim2\,n_0$, evident when comparing $P_{2 n_0}=13.4$-$30.4\MeVfm3$ from Ref.~\citep{2021PhRvD.104f3003L} with the present $P_{2 n_0}=22.0$-$27.1\MeVfm3$, both at the $1\sigma$ credible level, in the left panel of the NS case. 

As shown in Table~\ref{tb:comparison1}, the $1\sigma$ credible intervals of HSs are much larger than those of pure NSs. Similar results for $R_{1.4}$ (around $12\km$) are found for both pure NSs and HSs, as revealed by a consistent framework~\citep{2018ApJ...862...98Z} for dealing with single nucleons and nuclear many-body systems from the quark level. Many previous combined analyses of multimessenger observations also support this consistency (e.g., in~\citep{2020NatAs...4..625C}). However, there remains relatively large uncertainty for $M_{\rm TOV}$ in the range of $\sim2.0$-$3.3\Msun$.

From Fig. \ref{fig:EOS params pdf}, we learn that $K_0$ is more sensitive to the XMM-Newton data, while $L_0$ is more sensitive to both the XMM-Newton data and the HESS data. The HESS data has a significant modulation effect on $L_0$ (correspondingly to $R_{1.4}$ and $\Lambda_{1.4}$ as shown in Fig.~\ref{fig:star params pdf}) in both cases of pure NSs and HSs, as well as on $n_{\rm t}$ for HSs. 
While the saturation property $L_0$ benefits from such compensation, yielding a value of $41.7$ ($38.4\MeV$) in the case of pure NS (HS), a conflict arises for $n_{\rm t}$ in the range of $1.08$-$3.98\,n_0$ at the $1\sigma$ credible level; Similarly, the predicted $1\sigma$ credible interval for the sound speed squared is possibly in the range of $0.39$-$0.97$ at high-density cores of NSs (See Table \ref{tb:comparison2}). 
The influence of individual mass and radius constraints on the phase transition parameters exhibits similarities to our previous investigations~\citep{2020ApJ...904..103M, 2021ApJ...913...27L}. 
See discussions also in Ref.~\citep{2022PhRvC.105c5808D}.
In particular, consistent predictions are obtained for the saturation properties $K_0$, $E_{\rm sym}^0$, and $m^*_0$ for the cases with or without strangeness phase transition, with slightly larger ranges for those in the case of HSs, as detailed in Table \ref{tb:comparison2}. 
When accounting for the possibility of a phase transition, the resulting $L_0$ is around $41.7\MeV$, close to the EoS~\citep{2018ApJ...862...98Z} from the quark level previously constructed in the QMF framework.

\begin{table*}[t]
    \vspace{-0.4cm}
  \centering
  \caption {Most probable intervals of various properties for nuclear matter saturation and phase transition (quoted as median+68\% credible interval). The priors of these parameters are also listed in the second row.
  }
    \setlength{\tabcolsep}{0.8pt}
\renewcommand\arraystretch{1.3}
\begin{ruledtabular}
\begin{tabular*}{\hsize}{@{}@{\extracolsep{\fill}}lcccccccc@{}}
     & $K_0$&$E_{\rm sym}^0$&$L_0$& $m^*_0/m_N$&$n_{\rm t}/n_0$&$\mu_{\rm crit}$&$\Delta\varepsilon/\varepsilon_{\rm t}$& $c_s^2$\\
       \hline  
        prior &U$(220,280)$ &U$(30,35)$ & U$(20,120)$ &U$(0.65,0.8)$  &U$(1,4)$  &/  &U$(0,2)$  & U$(1/3,1)$  \\ 
     \hline             
         NS (w.o. XMM) &$259.63_{-15.53}^{+13.92}$ &$32.64_{-1.77}^{+1.58}$  &$53.56_{-21.02}^{+23.05}$ &$0.77_{-0.02}^{+0.01}$ &/ &/ &/ &/ \\ 
        NS (with XMM) &$265.06_{-14.85}^{+10.70}$ &$32.62_{-1.65}^{+1.61}$  &$77.50_{-21.34}^{+10.90}$ &$0.76_{-0.02}^{+0.02}$ &/ &/ & / &/ \\    
        NS (plus HESS) &$265.11_{-15.81}^{+10.23}$ &$32.76_{-1.74}^{+1.51}$ &$38.39_{-11.41}^{+17.68}$ &$0.76_{-0.02}^{+0.02}$ &/ &/  &/ &/\\            
    \hline
    HS (w.o. XMM)  &$259.85_{-18.83}^{+14.01}$ &$32.66_{-1.79}^{+1.61}$  
    &$62.13_{-25.00}^{+20.44}$ 
    &$0.77_{-0.03}^{+0.02}$ 
    &$2.28_{-0.72}^{+0.92}$  
    &$1078.82_{-69.80}^{+125.22}$  &$0.38_{-0.24}^{+0.35}$  
    &$0.77_{-0.18}^{+0.16}$  \\ 
    HS (with XMM) &$265.86_{-17.86}^{+10.25}$ &$32.73_{-1.77}^{+1.52}$  
    &$76.60_{-23.76}^{+12.79}$ 
    &$0.75_{-0.02}^{+0.03}$ 
    &$3.11_{-1.05}^{+0.69}$ 
    &$1311.67_{-239.94}^{+1063.26}$ &$0.31_{-0.22}^{+0.95}$ & $0.68_{-0.21}^{+0.21}$  \\ 
    HS (plus HESS) &$260.99_{-19.61}^{+13.15}$ &$32.73_{-1.63}^{+1.51}$  
    &$41.74_{-13.27}^{+20.48}$ &$0.76_{-0.02}^{+0.03}$& $2.38_{-1.11}^{+1.21}$  &$1101.85_{-116.53}^{+1198.91}$  &$0.37_{-0.26}^{+0.65}$     &$0.68_{-0.21}^{+0.20}$  \\      
  \end{tabular*}
  \end{ruledtabular}
          \vspace{-0.2cm}
  \label{tb:comparison2}
\end{table*}

\section{Summary and perspective}
\label{sec:summary}

NSs provide a distinctive environment for exploring fundamental nuclear physics. It offers a unique opportunity to study EoS of cold matters in equilibrium, whose density is a few orders of magnitude higher than that of typical nuclear matters on Earth. 
The pulse-profile modeling method has been successfully applied to infer the mass and radius of NSs conditional on the thermal emission of hot spots on the stellar surface. 
The latest mass and radius inferences of PSR J0030+0451 and PSR J0740+6620 are obtained from the joint analysis of the NICER and XMM-Newton data. 
It provides a preliminary constraint on the EoS together with the observations of the supernova remnant HESS J1731-347 and gravitational waves of GW170817.
As expected, when incorporating the updated radius measurement of light PSR J0030+0451 integrating XMM-Newton data, the present analysis establishes a refined constraint on the stiffness at $\sim2\,n_0$.
In particular, to compromise the different results regarding the radius of light NSs around $1.4\Msun$ between the HESS J1731-347 data and the updated PSR J0030+0451 data, the previously unknown isospin-asymmetric contribution of the EoS can be better determined. 
For example, the symmetry energy slope tends to be a low value of around $40\MeV$, even with the inclusion of possible non-nucleon degrees of freedom.
It emphasizes the need for collaborative data analysis and methodological refinements.

In the current work, seven parameters collaboratively determine the EoS prior, each parameter carrying specific physical significance associated with the composition and phase state of NS matter.
This framework allows for a direct connection between NS observations and microphysics-driven investigations of the EoS across various density regimes and facilitates a connection with the ongoing research efforts in the field of relativistic HIC~\citep{2023PhRvD.107d3005L}.
Moreover, the versatility of this analysis allows for straightforward extensions to higher-dimensional models, accommodating the inclusion of additional particles within NSs, such as hyperons~\citep{2023ApJ...942...55S} and dark matter~\citep{2022ApJ...936...69M}, to contribute a more comprehensive understanding of the interplay between different phase states of NS matter.
While establishing a direct connection between QCD and NS EoS remains challenging, the current framework offers a pragmatic approach to translating NS observations into the EoS microphysics. 
This approach supports quantitative studies of the EoS at different density regimes, inferring the values of unresolved properties of nuclear saturation ($K_0, E_{\rm sym}^0, L_0, m_{0}^*$) and deconfinement phase transition ($n_{\rm t}/n_0, \Delta\varepsilon/\varepsilon_{\rm t}, c_s^2,\mu_{\rm crit}$), particularly with the increasing accumulation of important NS observables, such as mass and radius. 

Nevertheless, the uncertainties of the background estimation and the emission configurations of hot regions limit the accuracy of the inferred mass and radius of NSs. 
There is a possibility that the current radius measurement, contingent on the XMM data set, which provides an indirect NICER background constraint, may face potential advancements in the future~\citep{2021ApJ...918L..27R}.
Addressing non-thermal emission components, exemplified by cases like PSR J0437-4715, is crucial~\citep{2013ApJ...762...96B,2016MNRAS.463.2612G,2019MNRAS.490.5848G}. 
Consequently, achieving more accurate measurements requires the future x-ray telescopes with large effective area, high energy resolution, and high time resolution in the soft x-ray band, as well as adequate imaging capability to discriminate the background, e.g.\ the planned enhanced X-ray Timing and Polarimetry Mission (eXTP)~\cite{2019SCPMA..6229503W}, the Spectroscopic Time-Resolving Observatory for Broadband Energy X-rays (STROBE-X)~\citep{2019arXiv190303035R}, and the Advanced Telescope for High Energy Astrophysics (NewAthena)~\citep{2023SPIE12679E..02B}. 
The follow-up X-ray Telescope (FXT), one of the payloads aboard the recently launched Einstein Probe (EP) mission, has space resolution $\sim {30}^{\prime\prime}$ and time resolution $\sim$ 100\,$\mu$s~\citep{2023ExA....55..603C}, of which the former property is capable of providing accurate background estimation and the latter could help to perform the timing analysis of the CCO and obtain a better pulse profile if any. 
Currently, it is important to note that the systematics and the physical assumptions behind the measurements of the CCO in the supernova remnant HESS J1731-347, for example, are huge.
With expectations of improved measurements in the near future, ongoing endeavors aim to enhance our understanding of the inner structure of NSs~\cite{2022MNRAS.515.5071M,2024arXiv240114930A}.

\medskip
\acknowledgments
We are thankful to Profs. J. B. Wei, X. Zhou and the XMU neutron star group for helpful discussions. 
The work is supported by National SKA Program of China (No.~2020SKA0120300), National Natural Science Foundation of China (grant Nos.~12273028, 12373051 and 12333007), and International Partnership Program of Chinese Academy of Sciences (Grant No.113111KYSB20190020).

\bibliography{ref}
\bibliographystyle{apsrev4-1}  

\end{document}